\documentclass[10pt,conference]{IEEEtran}
\IEEEoverridecommandlockouts
\usepackage{cite}
\usepackage{amsmath,amssymb,amsfonts}
\usepackage{algorithmic}
\usepackage{graphicx}
\usepackage{textcomp}
\usepackage{enumitem}
\usepackage{xcolor}
\def\BibTeX{{\rm B\kern-.05em{\sc i\kern-.025em b}\kern-.08em
    T\kern-.1667em\lower.7ex\hbox{E}\kern-.125emX}}

\usepackage{graphicx}
\usepackage{lipsum}
\usepackage{pdfpages}
\usepackage{float}
\usepackage{subcaption} % subfigures
\usepackage{nicefrac}
\usepackage{url}
\usepackage{xcolor}

\usepackage{booktabs}
\usepackage{comment}
\usepackage{color,soul} % package to highlight text
\usepackage{array}
\usepackage{multirow}
\usepackage{bigstrut}
\usepackage{subcaption}
\usepackage{units}
\usepackage{flushend}
\usepackage{url}
\usepackage{fontawesome}
\usepackage{hyperref}
\usepackage{colortbl}

\usepackage{makecell}
\usepackage{tcolorbox}
\tcbuselibrary{breakable}

\usepackage[utf8]{inputenc}

\usepackage{fancyvrb} % breaking lines with verbatim
\usepackage{fvextra} % breaking lines with verbatim
\usepackage{microtype}

\usepackage{tabularx}

\usepackage[normalem]{ulem}
\usepackage{rotating}
\usepackage{color}
\usepackage{tabularray}

\usepackage{pdfpages}

\usepackage{tcolorbox}
\usepackage{listings}
\usepackage{fvextra}

\usepackage{tabularray, fontawesome}
\usepackage[normalem]{ulem}
\usepackage{rotating}
\usepackage{color, xcolor, float,soul}
\usepackage{array, pgfplots}

\usepackage{lipsum}
\usepackage{ulem}

\usepackage[utf8]{inputenc}   % for pdfLaTeX
\usepackage[T1]{fontenc}

\usepackage{tcolorbox}
\usetikzlibrary{tikzmark,calc}
\usepackage{calc}

\usepackage{colortbl}

\colorlet{slightdarkgrey}{black!57}
\tcbuselibrary{xparse}

\begin{document}

\newtcolorbox{gradientbox}[2][]{enhanced, colback=#2!50, colframe=#2, width=1.2em, height=1.2em, valign=center, halign=center, fontupper=\footnotesize\bfseries, #1}

\newcolumntype{P}[1]{>{\RaggedRight\hspace{0pt}}p{#1}}
\definecolor{tableGray}{RGB}{243, 244, 245}
\definecolor{borderGray}{RGB}{229, 230, 233}

\def\mybarhhigh#1#2{%%
   {\color{black}\rule{#1mm}{8pt}}  #2}
   
\def\mybar2#1#2#3#4#5{%%
  %#1s 
   {\color{black}\rule{4pt}{#1mm}}
 {\color{black}\rule{4pt}{#2mm}}
  {\color{black}\rule{4pt}{#3mm}}
   {\color{black}\rule{4pt}{#4mm}}
    {\color{black}\rule{4pt}{#5mm}}
   }

\newtcolorbox{boxA}{
    colback = tableGray, % background color
    boxrule = 0pt  % no borders
}

\def\blackbar#1#2{%%
   {\color{black}\rule{#1mm}{4pt}}  #2}

\def\verticalbar#1#2{%%
   {\color{black}\rule{4pt}{#1pt}}  #2}

\def\multicolorbar#1#2#3#4#5#6{%%
   {\color{black!20}\rule{#1mm}{6pt}}%
   {\color{black!40}\rule{#2mm}{6pt}}%
   {\color{black!60}\rule{#3mm}{6pt}}%
   {\color{black!75}\rule{#4mm}{6pt}}%
   {\color{black!100}\rule{#5mm}{6pt} #6}%
}

\def\shadedbar#1#2#3#4#5{%%
  \begingroup
  \setlength{\fboxsep}{0pt}% Remove padding around the text
  \def\heightbox{\raisebox{0pt}[6pt][0pt]{}}% Define a macro for the fixed height
  \colorbox{black!100}{\makebox[#1mm][c]{\heightbox{\scriptsize\textcolor{white}{#1}}}}%
  \colorbox{black!75}{\makebox[#2mm][c]{\heightbox{\scriptsize\textcolor{white}{#2}}}}%
  \colorbox{black!60}{\makebox[#3mm][c]{\heightbox{\scriptsize\textcolor{white}{#3}}}}%
  \colorbox{black!40}{\makebox[#4mm][c]{\heightbox{\scriptsize\textcolor{white}{#4}}}}%
  \colorbox{black!20}{\makebox[#5mm][c]{\heightbox{\scriptsize\textcolor{white}{#5}}}}%
  \endgroup
}

\definecolor{TekheletPurple}{HTML}{572F96}
\newcommand{\boldification}[1]{\ifdraft\indent **\textbf{#1}**\\\indent\else\relax\fi}
\newcommand{\pQuote}[2]{{\color{TekheletPurple}\textit{``#1''}~({#2})}}

\DefineVerbatimEnvironment{Verbatim}{Verbatim}{breaklines=true}

\newcolumntype{P}[1]{>{\RaggedRight\hspace{0pt}}p{#1}}
\definecolor{tableGray}{RGB}{243, 244, 245}
\definecolor{borderGray}{RGB}{229, 230, 233}

\def\blackbar#1#2{%%
   {\color{black}\rule{#1mm}{4pt}}  #2}

\def\verticalbar#1#2{%%
   {\color{black}\rule{4pt}{#1pt}}  #2}

\def\multicolorbar#1#2#3#4#5#6{%%
   {\color{black!20}\rule{#1mm}{6pt}}%
   {\color{black!40}\rule{#2mm}{6pt}}%
   {\color{black!60}\rule{#3mm}{6pt}}%
   {\color{black!75}\rule{#4mm}{6pt}}%
   {\color{black!100}\rule{#5mm}{6pt} #6}%
}

\newtcolorbox{takeawayBox}{
  breakable,
  enhanced,
  sharp corners,
  colback=TekheletPurple!20!white,             % Light gray background
  colframe=TekheletPurple!20!white,            
  boxrule=0.5pt,
  left=3pt,
  right=3pt,
  top=3pt,
  bottom=3pt,
  before skip=1pt,
  after skip=1pt,
  fonttitle=\bfseries\small,
  coltitle=black,
  attach title to upper={},
  separator sign={\ },
}
   
\def\shadedbar#1#2#3#4#5{%%
  \begingroup
  \setlength{\fboxsep}{0pt}% Remove padding around the text
  \def\heightbox{\raisebox{0pt}[6pt][0pt]{}}% Define a macro for the fixed height
  \colorbox{black!100}{\makebox[#1mm][c]{\heightbox{\scriptsize\textcolor{white}{#1}}}}%
  \colorbox{black!75}{\makebox[#2mm][c]{\heightbox{\scriptsize\textcolor{white}{#2}}}}%
  \colorbox{black!60}{\makebox[#3mm][c]{\heightbox{\scriptsize\textcolor{white}{#3}}}}%
  \colorbox{black!40}{\makebox[#4mm][c]{\heightbox{\scriptsize\textcolor{white}{#4}}}}%
  \colorbox{black!20}{\makebox[#5mm][c]{\heightbox{\scriptsize\textcolor{white}{#5}}}}%
  \endgroup
}

%
% \usepackage{mathptmx}      % use Times fonts if available on your TeX system
%
% insert here the call for the packages your document requires
%\usepackage{latexsym}
% etc.
%
% please place your own definitions here and don't use \def but
% \newcommand{}{}
%
% Insert the name of "your journal" with
% \journalname{myjournal}
%

% Command to create comments in the document

% \renewcommand{\arraystretch}{1.5}  % Adjust row height

% DEFINING THE HIGHLIGHT BOX
\newtcolorbox{highlightbox}{
	colback=gray!20, % Background color
	colframe=black,   % Border color
	boxsep=2pt,       % Padding around the box content
	arc=0pt,          % Corner radius
	boxrule=1pt       % Border thickness
}

\title{Developer Task Delegation to LLMs and the Trust Factor: A Mixed Methods Study}

\author{Samuel Ferino, Rashina Hoda, John Grundy, Christoph Treude, Hashini Gunatilake% <-this % stops a space
\IEEEcompsocitemizethanks{\IEEEcompsocthanksitem S. Ferino, R. Hoda, J. Grundy, H. Gunatilake are with the Faculty of Information Technology,
Monash University, Melbourne, Australia. E-mail: \{samuel.demouraferino, rashina.hoda, john.grundy, hashini.gunatilake\}@monash.edu \protect\\
\IEEEcompsocthanksitem C. Treude is with the School of Computing and Information Systems, Singapore Management University, Singapore. E-mail: ctreude@smu.sg\protect\\
}% <-this % stops a space
}
\maketitle

\begin{abstract}
 With the rise of Large Language Models (LLMs), software developers are presented with a powerful new approach: delegating software development tasks to LLMs. Existing studies have explored task delegation for a subset of tasks and factors. Critically, what tasks developers delegate to LLMs, to what extent, and why they delegate them, is not yet well understood. We explore these questions through 22 interviews and a survey with 160 software practitioners. Developers reported employing four modes of task delegation to LLMs: full delegation, partial delegation, self-delegation, and task avoidance. Drawing on our qualitative findings, we discuss the rationales behind these choices and explore how task difficulty, task risk, developer motivation, and trust appear to be associated with software development task delegation to LLMs. Our statistical analysis reveals that trust is a significant factor associated with task delegation to LLMs. Finally, we share actionable implications for human-AI collaboration in software research, and for software practitioners to make the most of their newly found powers of delegation.
\end{abstract}


\section{Introduction}

The rise of large language models (LLMs) has equipped software developers with an unprecedented ability to delegate development tasks to coding agents \cite{ozkaya:2022}. The 2025 Stack Overflow Developer Survey reported that 78.5\% of developers are now using AI tools in their development process \cite{stackoverflow:2025}. The 2025 DORA report \cite{dora:2025} found that developers frequently rely on AI for tasks such as writing new code, modifying existing code, creating test cases, and explaining concepts. Similar trends are also quoted in industry practice. For example, Microsoft CEO Satya Nadella reported that \textit{``up to 30\% of the company's code was written by AI''} \cite{zeff:2025}. However, many recent incidents, such as Replit-agent failure \cite{fortune:2025, techco:2025}, highlight the risks associated with excessive delegation to autonomous systems without adequate safeguards. In that case, the agent reportedly made\textit{``unauthorised changes to live infrastructure, wiping out data for more than 1200 executives and over 1,190 companies''}. These incidents underscore the importance of maintaining appropriate human oversight and control when integrating AI into software development workflows.
There are also concerns that delegating too many tasks to LLM agents ``dumbs-down'' human software developers and reduces their ability to review and correct LLM-made errors \cite{mohamed2026impact,ferino2:2026}.

While developers are pushing the limits of delegating tasks to LLMs in practice, with mixed consequences, research is beginning to explore the advantages and risks of using LLMs and the different ways in which they are being used for software development tasks. For example, LLMs' automation capabilities are gaining traction as a promising solution to assist developers by handling tedious and repetitive tasks \cite{hou:2024, zheng:2025, ferino:2026}, such as reading documentation and web searches \cite{meyer:2019}. In some cases, automation has been found to enable developers to shift their focus to other activities that demand their expertise and creativity \cite{jackson:2025, groeneveld:2021}. Automation of some tasks, such as code generation and information retrieval, has been found to reduce developer effort \cite{ferino2:2025}. Hassan et al. \cite{hassan:2026} proposed Software Engineering (SE) 3.0, an AI-native approach for development between human developers and AI teammates. However, drawbacks of adopting LLMs, such as fragmenting developers' mental model of the codebase and hindering their SE skills are also emerging \cite{ferino2:2025}.

% Based on this, %%Hassan did not propose SE3.0 based on what is in the previous sentence 

% , were also evident in the same study %%same here sounds like hassan 2026 which it is not

Although developers use LLMs mostly for coding-related tasks \cite{dora:2025}, recent research \cite{lambiase:2025, sergeyuk:2025, spitzer:2025, choudhuri:2025} explored some aspects of task delegation for specific development tasks (e.g., implementing new features and writing tests), as well as specific factors (e.g., perceived task value) influencing AI adoption and use. However, Lambiase et al. \cite{lambiase:2025} found  LLM adoption to be complex and potentially associated with multiple factors. While these studies have started to build an understanding of AI adoption and use in SE, to what extent and why developers delegate coding-related tasks to LLMs is not known \cite{choudhuri:2026}.
Motivated by the need to better understand how software developers in real-world practice are actually delegating tasks to LLMs, to what extent, and why, we conducted a mixed-methods empirical study. Our aim was to answer the following research questions (RQs):

\begin{itemize}
\item  \textbf{RQ1}. \textbf{\textit{How do software developers delegate tasks to LLMs? Which tasks and to what extent?}} 
% { 
% \begin{itemize}
    % \item \textit{Which tasks do they delegate to LLMs?}
    % \item \textit{To what extent do they delegate these tasks to LLMs?}
% \end{itemize} } 

\item  \textbf{RQ2}. \textbf{\textit{Why do software developers delegate certain tasks to LLMs?}}

\end{itemize}

To answer our RQs, we conducted a mixed-methods study involving semi-structured interviews with 22 software practitioners, followed by a survey focused on software development tasks with 160 software practitioners. From this study, we identified four developer task delegation to LLMs modes: full delegation, partial delegation, self-delegation, and task avoidance grounded in empirical evidence. \textit{Full delegation} involves minimal developer intervention in LLMs' actions, only reviewing LLMs' final output. \textit{Partial delegation} requires active, close management of LLMs' actions, demanding higher developer engagement. \textit{Self-delegation} occurs when developers bypass LLMs entirely, relying instead on traditional resources like Google search or Stack Overflow. \textit{Task avoidance} occurs when developers choose to completely evade a task, despite having access to LLM assistance. We identified that \textit{task difficulty}, \textit{task risk}, \textit{developer motivation}, and \textit{developer trust towards LLMs} are key factors shaping task delegation. From our statistical analysis, we found evidence that \textit{developer trust towards LLMs is a significant factor} associated with full or partial task delegation.

The key  contributions of our research are: 
\begin{itemize}
    % \item We carried out a mixed-methods study with 22 interviews and 160 survey participants to better understand software developer task delegation to LLMs in practice;
    \item We identified four modes of delegation used and reasons for use;
    \item We identified the main factors that seem to be shaping human software developer choices to delegate or partially delegate tasks to LLMs.
    % \item We identified a set of implications for current practice and future research.
\end{itemize}

The rest of the paper is structured as follows: section \ref{sec:methodology} describes our mixed method study methodology, and section \ref{sec:results} presents our findings. Section \ref{sec:relatedWork} summarises key related works and section \ref{sec:discussion} presents a set of implications for practice and future research. Section \ref{sec:threatsToValidity} presents the study limitations and section \ref{sec:conclusion} summarises our study  conclusions. 

% \textcolor{red}{fix the order now that we moved RW to before discussion}

\begin{figure}
    \centering
    \includegraphics[width=\linewidth]{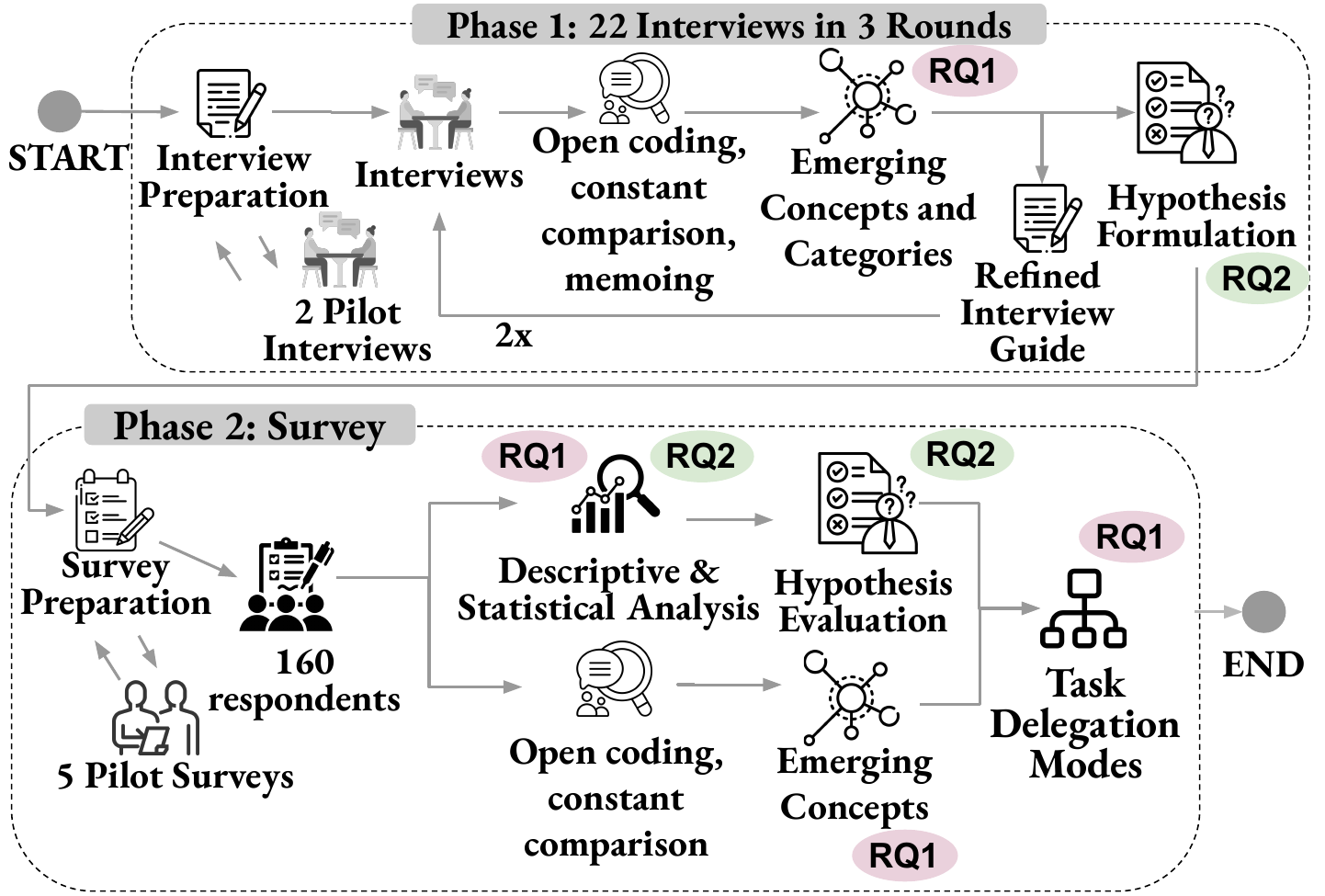}
    \caption{Overview of the mixed methods research.}
    \label{fig:methodology}
\end{figure}
\section{Study Design} \label{sec:methodology}

Recent studies explore how LLMs influence software practitioners' task delegation decisions (e.g., \cite{lambiase:2025,choudhuri:2025,sergeyuk:2025}) and are summarised in section \ref{sec:relatedWork}. To the best of our knowledge, this is the first study to present four modes of developer task delegation to LLMs. These modes were identified through the analysis of exploratory interviews and subsequently complemented by a survey. In this paper, we also investigate the socio-technical factors behind these delegation decisions, examining how task difficulty, task risk, developer motivation, and trust in LLMs are associated with the mode of task delegation.

To do this, we applied a mixed methods research approach using an \textit{exploratory sequential design} \cite{storey:2025}. We first explored the use of LLMs in software development through qualitative semi-structured interviews. We then followed these interviews with practitioner mixed-data survey, with \textit{quantitative} data from the closed-ended questions and \textit{qualitative} data from the open-ended questions. Figure \ref{fig:methodology} provides an overview of our study methodology. The supplementary material \cite{researchArtifact} includes the pre-interview questionnaire, the interview guide, the survey questions, examples of qualitative data analysis, and quantitative data analysis scripts, in line with the requirements of the ethical research approval gained for the study. In the following, we describe the two sequential phases of the study. The interviews helped to address RQ1, while the hypotheses resulting from the interviews and the survey results helped to address some of RQ1 and all of RQ2.

% , anonymised survey data

\subsection{Phase 1: Semi-Structured Interviews}

\subsubsection{Interview Data Collection} 

% Aligned with  Gunatilake et al. \cite{gunatilake:2023}, 
We recruited software developers with experience in using LLMs for software development from LinkedIn, X (formerly Twitter), personal connections, and through word-of-mouth snowballing. We use the term software developer \textit{``as an umbrella term in the SE context''},  encompassing a range of roles including but not limited to web development and requirements engineering \cite{gunatilake:2023}. Each interviewee filled a 10-minute pre-interview questionnaire and participated in a semi-structured interview. 
The pre-interview questionnaire collected demographics, LLM experience, and \textit{Attitudes towards Artificial Intelligence} (ATAI scale) \cite{sindermann:2021}. ATAI scale involves the following five eleven-point scale questions: (1) \textit{I fear artificial intelligence}; (2) \textit{I trust AI}; (3) \textit{AI will destroy humankind;} (4) \textit{AI will benefit humankind}; and (5) \textit{AI will cause many job losses}. It was used to understand participants' attitudes towards AI in general, not limited to LLMs. 

Interviews were then conducted by the first author over three rounds between October (2024) and September (2025) and lasted between 34 to 58 minutes. Although not mentioned in the study advertisement to avoid unscrupulous interviewees, we provided a \textit{(currency and amount withheld for anonymous review)} gift voucher to each interview participant after the interview to express our gratitude for their commitment.
Our semi-structured interview questions explored participants' experiences, perceptions of adopting LLMs in software development, factors associated with their adoption, and their attitudes regarding task delegation to LLMs. We conducted two pilot interviews with experienced software practitioners to evaluate the interview questions and structure, which resulted in minor refinements. We excluded the data from these pilot interviews from the final interview pool, as they did not have experience using LLMs for software development.

We used Qualtrics for the pre-interview questionnaire and Zoom for recording and transcription of the interviews. Interviewees varied in years and level of experience, geographical locations, and domains. Table \ref{tab:demographicInfoInterview} presents this demographic information including their experience using LLMs for software development. 
% Portuguese

\begin{figure*}
    \centering
    \includegraphics[width=\textwidth]{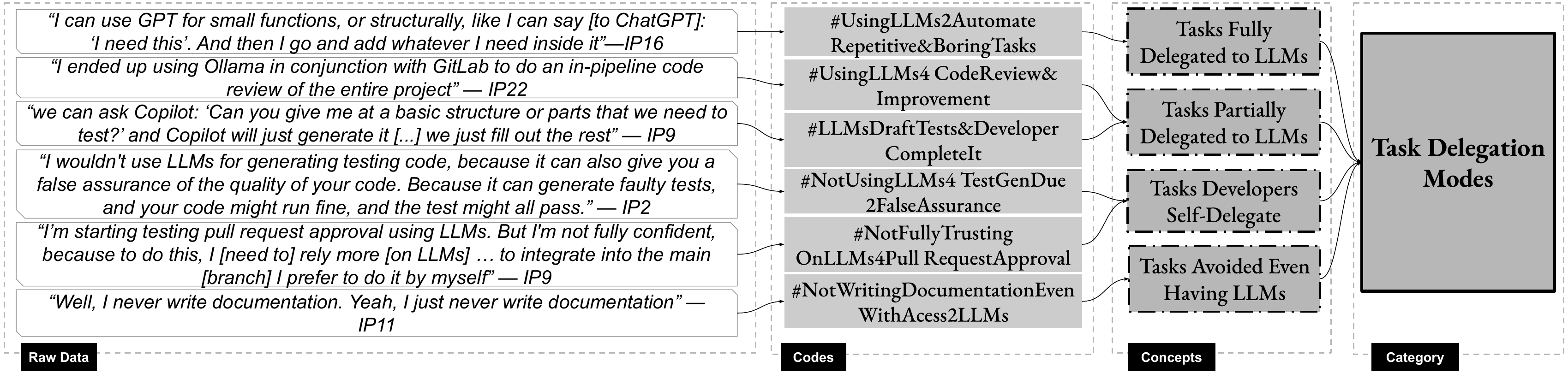}
    \caption{Emergence of the category ``\textsl{Task Delegation Modes}'' from \textit{raw data} → \textit{codes} → \textit{concepts} → \textit{category} through constant comparison. }
    \label{fig:OpenCodeConstantComparison}
\end{figure*}

\subsubsection{Interview Data Analysis} 
We used socio-technical grounded theory \textit{for data analysis} (STGT4DA) \cite{hoda:2024} to analyse the qualitative data from the semi-structured interviews to derive layered and nuanced findings, as has been applied in similar contexts \cite{madampe:2022, pant:2024, lambiase:2024}. The first author conducted the data analysis, applying open coding, constant comparison (see Figure \ref{fig:OpenCodeConstantComparison}), and memoing. Instead of applying quantitative metrics such as inter-rater/coder reliability, which are widely considered epistemologically incompatible with the inductive qualitative data analysis techniques~\cite{hoda:2024, oconner:2020, braun:2013, hollway:2012}, we focused on collaborative alignment~\cite{hoda:2024}: the co-authors reviewed the emerging codes and concepts regularly throughout the process, and we also conducted dedicated coding workshops to align our initial coding principles and critically review the concepts, categories, and socio-technical findings.

% Following the guidelines suggested by Hoda~\cite{hoda:2024}, who argues that formal quantitative coder metrics — such as inter-rater reliability and inter-coder reliability — are methodologically incompatible with inductive STGT4DA applications, we focused instead on interpretive validity through collaborative alignment. The other authors reviewed the emerging codes and concepts regularly throughout the process.  We also conducted dedicated STGT workshops to align our initial coding principles and critically review the concepts, categories, and socio-technical findings. 

% Instead of applying quantitative metrics such as inter-rater/coder reliability which are widely considered epistemologically incompatible with the inductive qualitative data analysis techniques [26, X, Y, Z], we focused collaborative alignment [Chapter 10, 26]: the other authors reviewed the emerging codes and concepts regularly throughout the process. We also conducted dedicated coding workshops to align our initial coding principles and critically review the concepts and categories and socio-technical findings. A sample memo emerged from the interview analysis is presented below.

Through the analysis of the interview data, we identified four {task delegation modes}  that our interviewees had adopted for tasks across the software development life cycle (SDLC):

\begin{itemize}
    \item \textit{fully delegating the task \faHourglassO} to LLMs, 
    \item \textit{partially delegating the task \faHourglass[3]} to LLMs, 
    \item \textit{self-delegation of tasks \faHourglass}, and
    \item \textit{task avoidance \faClose} even when LLMs were available. 
\end{itemize}

\label{sec:differentApproachesTowardsSETasks}

% participants (IP) have different attitudes regarding the use of LLMs for SE tasks. P

% Phase 2: 
\subsubsection{Hypotheses Formulation} \label{subsecsec:hypothesis}

Based on our analysis of the interview data, we also identified several potential factors associated with task delegation that emerged during the memoing process, as reported by participants: task difficulty, task risk, developer motivation, and developer trust towards LLMs.
Table \ref{tab:hypotheses} presents our hypotheses about the relationships between these potential factors and software developer task delegation behaviour. We define each emerging factor, grounding our insights in illustrative quotes from the interview participants in section \ref{factors}. A sample memo on one of the hypotheses is presented below. The supplementary package includes all memos supporting these hypotheses. 

% \textcolor{red}{For the coding, you currently have the coding evidence figure AND the memo but for the hypotheses development, there is nothing directly inside the paper. So, it would be better to replace the current memo with one that shows how one of the hypothesis is supported. You can share some options with me in email if you need help selecting which one. Otherwise the hypothesis development is the weakest part.} 

\begin{table}
\centering
\caption{Four hypotheses formulated from Interview data analysis.}
\label{tab:hypotheses}
\begin{tblr}{
  row{1} = {font=\bfseries},
    colspec = {Q[200]Q[408]},
  vline{1,3} = {-}{},
  hline{1,6} = {-}{0.08em},
  hline{2} = {-}{0.05em},
}
Hypothesis                                              & Grounding                                                                                      \\
$H_1$:~Difficulty $\rightarrow$ Delegation Mode~$(-)$ & Higher perceived task difficulty decreases delegation to LLMs                                  \\
$H_2$:~Risk $\rightarrow$ Delegation Mode~$(-)$  & Higher perceived risk of delegating a task reduces willingness to fully delegate               \\
$H_3$:~Motivation $\rightarrow$ Delegation Mode~$(-)$      & Higher intrinsic motivation to perform a task oneself reduces delegation                       \\
$H_4$:~Trust $\rightarrow$ Delegation Mode~$(+)$      & Higher trust in LLM capability for a specific task increases willingness to delegate that task 
\end{tblr}
\end{table}

\begin{tcolorbox}[float=htpb!,colback=gray!5!white,colframe=gray!75!black,title={\textbf{\footnotesize{Memo on \textit{``Trust $\rightarrow$ Delegation Mode~$(+)$''}}}}]
{\footnotesize  Trust is associated with task delegation in different situations. For instance, interview participants IP8 and IP12 describe vibe coding  - closely related to full delegation - as associated with \texttt{\#OverTrustingInLLMs} and having no safeguards. In this context, IP2, IP3, IP5, IP7, IP8, IP13, IP14, IP15, IP16, IP17, IP18, and IP19 suggest \texttt{\#ModerateTrustInLLMs} and \texttt{\#MandatoryLLMOutputReview\&Comphrension}, requiring more developer engagement when delegating tasks to LLMs. At the same time,  IP2 \texttt{\#DistrustsLLMs4EmergentTopics} based on previous experience, chose to self-delegate these tasks.}
\end{tcolorbox}

% {\footnotesize  Trust is associated with task delegation in different situations. For instance, interview participants IP8 and IP12 associate vibe coding - closely related to full delegation - with overtrust in LLMs, having no safeguards. Interview participants IP12 and IP22 delegate tasks according to their level of trust. IP12 mentions delegating unit and integration testing to LLMs because of trust in their ability to perform those tasks. At the same time, interview participant IP2 self-delegates tasks involving emergent topics due to distrust of LLMs' capabilities in these contexts based on previous experience.}
% especially the tasks developers decide to avoid

The findings from the interviews, including the hypotheses, motivated us to focus on better understanding task delegation in the next phase. For this, we decided to narrow down our focus to coding related-tasks, mainly because these are the tasks developers mostly delegate to the LLMs~\cite{dora:2025}. 

% \textcolor{red}{this short para may fit better here as a closing to Phase 1 and pre-cursor to Phase 2} 

\subsection{Phase 2: Survey}

% we conducted an anonymous 15-minute survey with software practitioners. Our survey followed Kitchenham and Pfleeger's guidelines \cite{kitchenham:2002}, combining both open and closed ended questions, including demographics, LLM experience, ATAI scale, task delegation modes, and Task Delegability to AI framework (TDAI). Aligned with our emerging factors, we employed a slightly modified version of the TDAI framework  \cite{lubars:2019, cvetkovic:2022} that includes LLMs and GenAI in the questions to assess our hypotheses, examining task delegation to LLM tools in SE. There are fourteen questions measured using a five-point Likert framework, which are organised across four aspects: \textit{difficulty}, \textit{risk}, \textit{motivation}, and \textit{trust}. We share the survey questions in the supplementary package \cite{researchArtifact}. Through five pilots, we narrowed down the long list of SE tasks developers use LLMs documented by Hou et al. \cite{hou:2024} to the following 17 coding-related tasks to fit our short survey format: \textit{code generation, code completion, code comment generation, code representation, API document smells, API recommendation, API document augment, API entity and relation extraction, type inference, code summarisation, code search, code understanding, code translation, method name generation, control flow graph generation, agile story point estimation}, and \textit{data analysis}. We decided to include the last two pilot survey responses into the survey dataset since the core questions did not change significantly after that.
% 

To better understand how our proposed task delegation modes manifest across a broader population and to test our hypotheses, 
we conducted an anonymous 15-minute survey with software practitioners. Our survey followed Kitchenham and Pfleeger's guidelines \cite{kitchenham:2002}, combining both open and closed ended questions, including demographics, LLM experience, ATAI scale, task delegation modes, and Task Delegability to AI framework (TDAI). Aligned with our emerging factors from the interview analysis, we employed a slightly modified version of the TDAI framework  \cite{lubars:2019, cvetkovic:2022} that includes LLMs and GenAI in the questions to assess our hypotheses regarding \textit{difficulty}, \textit{risk}, \textit{motivation}, and \textit{trust}. TDAI allowed us to measure our emerging factors through the survey, comprehending fourteen questions measured using a five-point Likert framework. We share the survey questions in the supplementary package \cite{researchArtifact}. Through five pilots, we narrowed down the long list of SE tasks developers use LLMs documented by Hou et al. \cite{hou:2024} to the following 17 coding-related tasks to fit our short survey format: \textit{code generation, code completion, code comment generation, code representation, API document smells, API recommendation, API document augment, API entity and relation extraction, type inference, code summarisation, code search, code understanding, code translation, method name generation, control flow graph generation, agile story point estimation}, and \textit{data analysis}. We decided to include the last two pilot survey responses into the survey dataset since the core questions did not change significantly after that.

%  X factors emerged from the interview analysis, and we focused on these 4 as they were common to the TDAI framework and allowed us to measure them through the survey

% \textcolor{red}{how does the existing TDAI match exactly with the 4 factors you 'discovered' from the interviews? One approach can be to say X factors emerged from the interview analysis, and we focused on these 4 as they were common to the TDAI framework and allowed us to measure them through the survey}

\subsubsection{Survey Data Collection} \label{subsubsec:measurementeInstrument}

We initially distributed our Qualtrics survey via LinkedIn and across our personal networks in December (2025).  Similar to the interview participants, the survey participants were required to have experience of using LLM tools for software development tasks. After receiving only one response over almost one month, we then distributed the survey via \textit{Prolific}, an online recruitment platform \cite{palan:2018}. To ensure the quality of the responses, we employed attention checks \cite{oppenheimer:2009} and authentication checks to filter AI-generated responses. We also leveraged built-in options in the Prolific platform to filter Prolific users based on IT-related job roles, such as software and cloud engineers. The survey was administered in January (2026) and then in May (2026). From the 187 responses collected via Prolific, our manual review resulted in 155 (82.8\%) accepted survey submissions, including only participants reporting having experience using LLMs for SE tasks. Each Prolific survey participant was rewarded \textit{(currency and amount withheld for anonymous review)} after completing the survey.  In total, we collected 160 survey responses, including 155 Prolific responses, three responses from LinkedIn, and two from the pilot survey.

% \textcolor{red}{quick explaination of why others were rejected?} including only participants with experience using LLMs for SE tasks.

\subsubsection{Survey Data Analysis} \label{subsubsec:plsSem}

To analyse the qualitative data from the survey collected through the responses to the open-ended questions, we applied STGT4DA, using open coding and constant comparison, similar to how we analysed the qualitative data from the interviews (see Figure \ref{fig:OpenCodeConstantComparison}).

To analyse the survey's quantitative data, we calculated descriptive statistics for each closed-ended survey item to identify overall response patterns and trends. We then performed statistical analysis to examine whether \textit{difficulty}, \textit{risk}, \textit{motivation}, and \textit{trust} are positively or negatively associated with task delegation to LLMs. Before conducting our statistical analysis, we assessed the psychometric properties of the TDAI framework through multilevel coefficient omega ($\omega$), which accounts for the nested structure of our data (repeated task ratings within participants), and variance inflation factor (VIF). Between-person reliability was strong across all subscales ($\omega_{between}$=0.744--0.947). Within-person reliability was acceptable for \textit{motivation} ($\omega_{within}$=0.785) and \textit{risk} ($\omega_{within}$=0.628), but modest for \textit{difficulty} ($\omega_{within}$=0.544) and \textit{trust} ($\omega_{within}$=0.512). Intraclass correlation coefficients (ICC=0.476--0.692) further confirmed that a substantial proportion of variance in all subscales reflected stable individual differences rather than task-specific responses. VIF values ranged 1.02--1.32 at the within-person level and 1.14--1.45 at the between-person level. Pairwise correlations did not exceed 0.47 at either level.

% We present the results in the supplemented package \cite{researchArtifact}. 

\begin{table}[h]
\caption{\small Interview Participant Demographics \footnotesize{(Dev: Developer; Eng: Engineer; XT: Total Professional Software Development Experience; XTL: LLMs for SE Experience; CLIP: Company's LLM Policy)}.}
\label{tab:demographicInfoInterview}
\resizebox{\columnwidth}{!}{%
\begin{tabular}{@{}llll@{}}
\toprule
\textbf{Location} & \textbf{\# of Participants} & \textbf{Role} & \textbf{\# of Participants} \\ \midrule
Australasia     & \mybarhhigh{33}{10}                          & Software Eng.           & \mybarhhigh{30.6}{6}                          \\
North America              & \mybarhhigh{22}{5}                          & Software Dev.        & \mybarhhigh{27}{5}                          \\
Latin America       & \mybarhhigh{18.18}{4}                          & Data Analyst   & \mybarhhigh{9}{2}                          \\
Asia            & \mybarhhigh{9}{2}                          &  AI Engineer            & \mybarhhigh{4.5}{1}                          \\
Europe     & \mybarhhigh{4.5}{1}                        & Others        & \mybarhhigh{32}{8}                          \\
% \cmidrule(r){1-2}
\midrule
\textbf{Domain}   & \textbf{\# of Participants} &  \textbf{Domain}   & \textbf{\# of Participants}                          \\ \midrule %\cmidrule(r){1-2}
IT              & \mybarhhigh{35}{13}                         & Others    & \mybarhhigh{33}{7}                           \\
Finance            & \mybarhhigh{9}{2}                        &          &                           \\
\midrule
\textbf{XT}       & \textbf{\# of Years}        & \textbf{XTL}  & \textbf{\# of Years}        \\ \midrule
<= 2 years            & \mybarhhigh{33}{10}                     &  < 1 year       & \mybarhhigh{35}{12}                           \\
3-5 years           & \mybarhhigh{31}{7}                          & 1-2 years    & \mybarhhigh{32}{8}                          \\
6-10 years           & \mybarhhigh{13}{3}                          & +2 years   & \mybarhhigh{9}{2}                          \\
+10 years              & \mybarhhigh{9}{2}                        &     &                         \\ 
\midrule
\textbf{LLM tools}       & \textbf{\# of Participants}        & \textbf{CLIP}  & \textbf{\# of Participants}        \\ \midrule
ChatGPT           & \mybarhhigh{18}{18}                          & Allow    & \mybarhhigh{37}{16}                          \\
Copilot              & \mybarhhigh{9}{7}                        &  No policy   &   \mybarhhigh{13}{3}                      \\ 
Claude              & \mybarhhigh{9}{7}                        &   Prohibit  &   \mybarhhigh{4.5}{1}                      \\ 
Others              & \mybarhhigh{9}{<5}                        & Others     &      \mybarhhigh{9}{2}                   \\ \bottomrule

\end{tabular}%
}
\end{table}

\begin{table}[h]
\caption{\small Survey Participants Demographics.}
\label{tab:demographicInfoSurvey}
\resizebox{\columnwidth}{!}{%
\begin{tabular}{@{}llll@{}}
\toprule
\textbf{Location} & \textbf{\# of Participants} & \textbf{Role} & \textbf{\# of Participants} \\ \midrule
Europe     & \mybarhhigh{25}{66}                          & Software Eng.           & \mybarhhigh{30.6}{49}                          \\
North America              & \mybarhhigh{17}{35}                          & Software Dev.        & \mybarhhigh{30.6}{49}                          \\
Asia            & \mybarhhigh{13.1}{23}                          & Web Dev.            & \mybarhhigh{7.5}{12}                          \\
Latin America       & \mybarhhigh{12.5}{20}                          & Frontend Dev.   & \mybarhhigh{5}{8}                          \\
Australasia     & \mybarhhigh{5}{8}                          & Cloud Eng.        & \mybarhhigh{3.1}{5}                          \\
Africa            & \mybarhhigh{5}{8}                           & Other  & \mybarhhigh{26.25}{37}                           \\ \midrule %\cmidrule(r){1-2}
\textbf{Domain}   & \textbf{\# of Participants} &  \textbf{Domain}   & \textbf{\# of Participants} \\ \midrule %\cmidrule(r){1-2}
IT              & \mybarhhigh{30}{92}                         & Healthcare    & \mybarhhigh{5.6}{9}                           \\
Finance            & \mybarhhigh{7.5}{12}                        & Manufacturing         & \mybarhhigh{5.6}{9}                          \\
Education    & \mybarhhigh{6.8}{11}                           &   Others             &    \mybarhhigh{16.8}{27}                          \\ \midrule
\textbf{XT}       & \textbf{\# of Years}        & \textbf{XTL}  & \textbf{\# of Years}        \\ \midrule
<= 2 years            & \mybarhhigh{23.1}{37}                     &  < 1 year       & \mybarhhigh{16.8}{28}                           \\
3-5 years           & \mybarhhigh{30}{53}                          & 1-2 years    & \mybarhhigh{35}{65}                          \\
6-10 years           & \mybarhhigh{27}{40}                          & 2-3 years   & \mybarhhigh{23.1}{36}                          \\
+10 years              & \mybarhhigh{18}{30}                        & +3 years    & \mybarhhigh{18}{31}                        \\ 

\midrule
\textbf{LLM tools}       & \textbf{\# of Participants}        & \textbf{CLIP}  & \textbf{\# of Participants}        \\ \midrule
ChatGPT           & \mybarhhigh{35}{139}                          & Restrict    & \mybarhhigh{37}{69}                          \\
Gemini              & \mybarhhigh{31}{98}                        &  Allow    &   \mybarhhigh{28}{51}                      \\ 
Claude              & \mybarhhigh{30}{92}                        &   No policy   &   \mybarhhigh{17}{29}                      \\ 
GitHub Copilot              & \mybarhhigh{25}{<89}                        & Prohibit     &      \mybarhhigh{4.5}{6}                   \\ 
Others              & \mybarhhigh{20}{<78}                        & Others     &      \mybarhhigh{3.6}{5}                   \\ \bottomrule
\end{tabular}%
}
\end{table}

We fitted a sequential binary mixed-effects logistic regression model \cite{tutz:1991} by decomposing the ordinal three task delegation modes outcome (i.e., \textit{full delegation}, \textit{partial delegation}, and \textit{self-delegation}) with sufficient data into two distinct, nested binary transitions. \textit{Task avoidance} was excluded from the regression model because of it is marginally selected in practice, having a small sample size (n=21). A standard ordinal logistic regression model would treat the outcome as a single linear progression and impose the proportional odds assumption, requiring each factor to have the same effect at every step of the scale. Preliminary testing showed that this assumption was violated, meaning that the relationship between the factors and task delegation mode changes depending on which transition is being modelled. Sequential binary logistic regression analysis addresses this issue by explicitly decomposing the ordinal outcome into two binary decisions that reflect the natural decision process underlying the data. 
We model   two distinct hierarchical decision stages, each estimated using a separate generalised linear mixed model with random intercepts for participant identity to account for repeated task ratings nested within participants:

\begin{figure*}
    \centering
    
    \includegraphics[width=\linewidth]{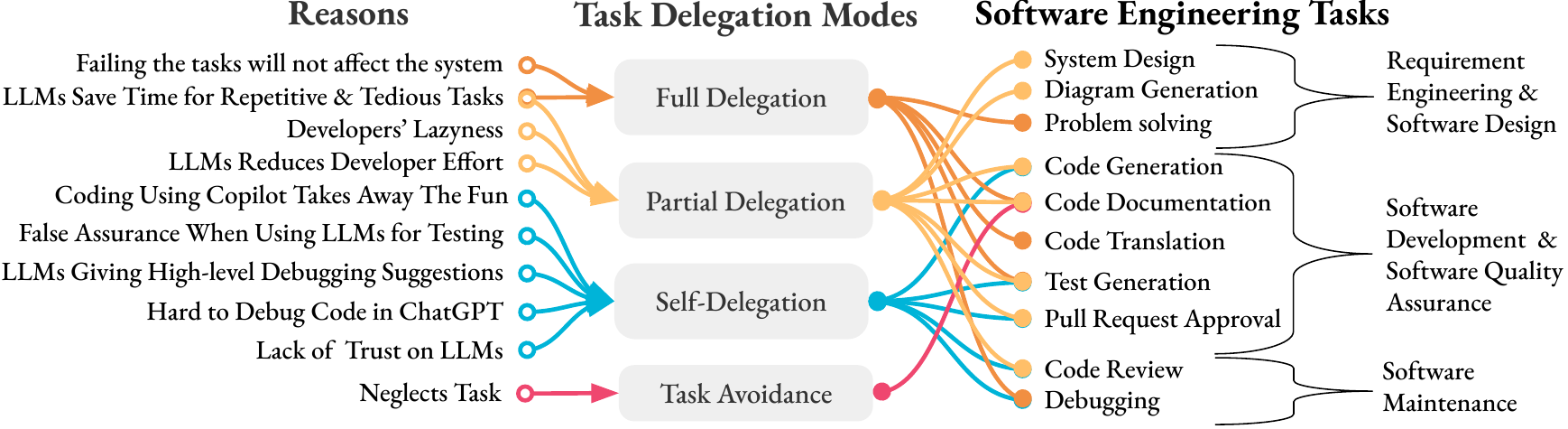}
    \caption{Task delegation modes \footnotesize{(Interview findings)}.}
    \label{fig:interviewFindings}
\end{figure*}

\begin{enumerate}
    \item \emph{Adoption stage} ($n = 857$):  It estimates the probability that a developer moves from a traditional developer-driven workflow to any level of LLM-assisted work. It includes the following task delegation modes: self-delegation ($n = 132$) and delegation (partial delegation [$n = 516$] + full delegation [$n = 209$]).
    \item \emph{Autonomy stage} ($n = 725$): It is conditional on LLM adoption and estimates the probability of transitioning from a supervised, human-in-the-loop workflow to complete delegation. It includes the following task delegation modes: partial delegation ($n = 516$) and full delegation ($n = 209$).
\end{enumerate}

\section{Findings}  \label{sec:results} 
\subsection{Participants}

Table \ref{tab:demographicInfoInterview} summarises the key demographics of the 22 interview study participants. Table \ref{tab:demographicInfoSurvey} summarises the key demographics of the 160 survey participants. The overall ATAI scale acceptance score was 6.0 (interviews) and 6.81 (survey), meaning participants have balanced attitudes towards AI.

\begin{figure}
    \centering
    \includegraphics[width=\linewidth]{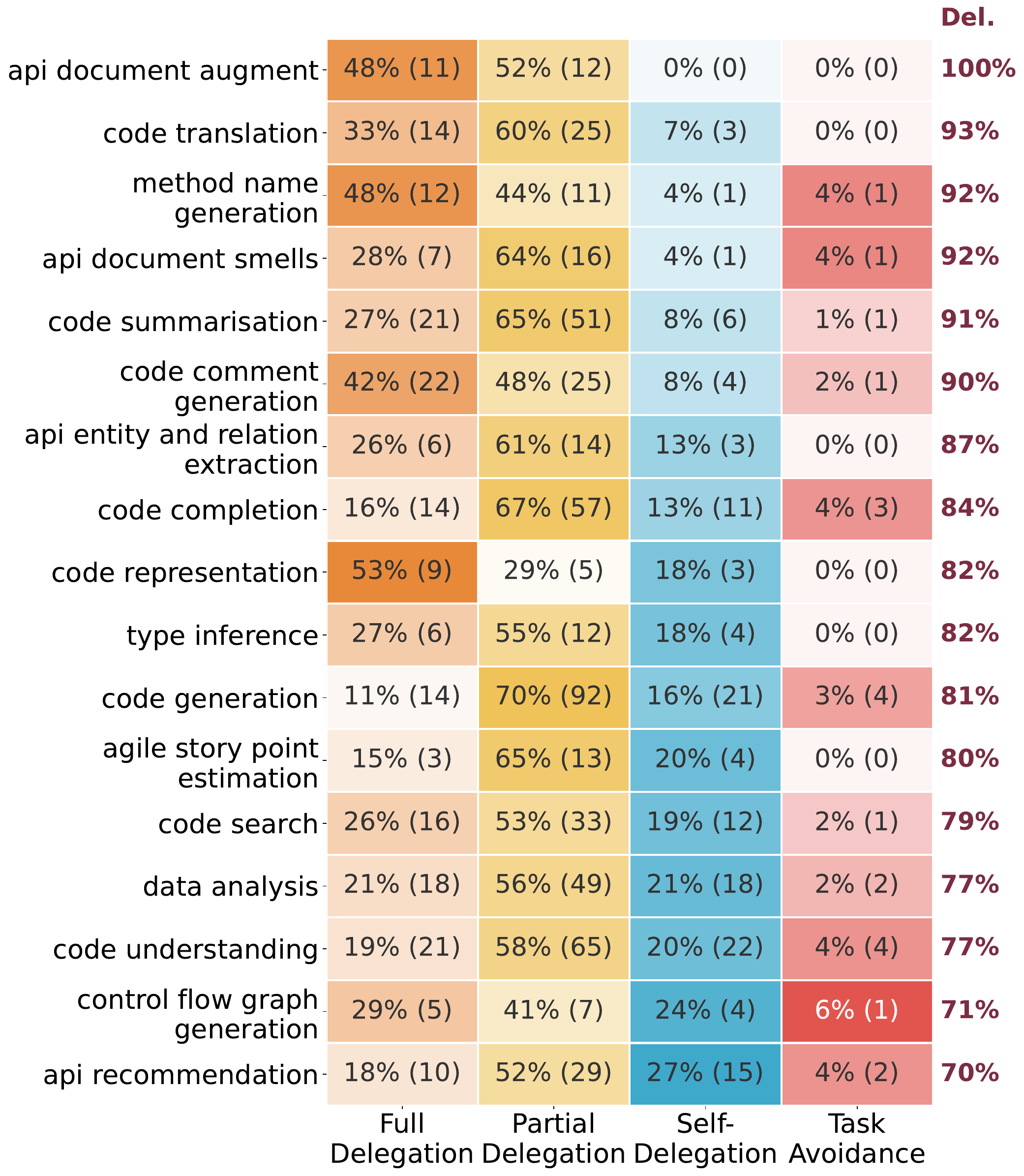}
    \caption{Distribution of task delegation modes  (\footnotesize{Survey findings; Del.: Fully and Partially Delegated}). }
    \label{fig:workingMode}
\end{figure}

%Total counts:
%AT:21, CT:137, PD: 536, FD:217

\subsection{Task Delegation with LLMs (RQ1)}
Figure \ref{fig:interviewFindings} summarises the SE tasks, task delegation modes, and reasons identified in the interviews. Our qualitative analysis reveals that participants' willingness to delegate tasks to LLMs is primary governed by opportunities for time savings, particularly for low-stakes tasks, and by situations in which they experience low personal motivation to perform the work themselves. Conversely, interview participants reported a strong preference for retaining manual control over tasks they found enjoyable, tasks for which they lacked trust in LLM capabilities, and tasks that exceeded the perceived technical limitations of current LLMs.

Figure \ref{fig:workingMode} illustrates how participants distribute their task delegation modes \textit{(full delegation, partial delegation, self-delegation, task avoidance)} across 17 distinct SE tasks. Across almost all tasks, \textit{partial delegation} to LLMs (n=516) emerged as the dominant  task delegation mode, indicating that developers predominantly use LLMs as collaborative tools rather than autonomous agents. While \textit{full delegation} (n=209) was observed for every task, it remained uncommon compared to partial delegation, even for tasks strongly associated with generative AI, such as code generation (n=15). 
In total, the \textit{self-delegation} mode (n=132) was the third most common delegation mode, occurring less frequently than both partial delegation and full delegation across many tasks. In contrast, the \textit{task avoidance}  mode was consistently the least common across tasks, indicating that participants rarely chose to forgo a task despite having access to LLM support. This pattern suggests that developers are generally willing to leverage LLM assistance to complete tasks rather than performing them entirely on their own, while outright task avoidance remains uncommon. However, we did not collect data on how frequently participants performed each task daily. Hence, differences in delegation patterns should be interpreted cautiously as they may reflect not only task characteristics but also variation in participants' exposure to or experience with particular tasks.

% \textcolor{red}{really? I thought we did..hmm}

Below we provide representative participant quotes in which our analysis is grounded. Survey data is used to supplement the interview findings for tasks that were included in the survey. In order to protect the privacy of our participants, we refer to the interview participants as IP1-IP22 and survey participants as SP1-SP160.

\subsubsection{\textbf{{Tasks developers fully or partially delegate}}} \label{sec:qualitativeFindings}

We found interview participants  \, \textit{partially \faHourglass[3]} \, or \textit{fully \faHourglassO} \,  delegating the following SE tasks to LLMs: system design, diagram generation, code generation, code translation, test generation, document generation, and code review. Their reasons for delegation are mainly that LLMs can free developers' time from low-stakes, repetitive or tedious tasks, and developers' self-admitted laziness. IP16 and IP13 express how they become dependent on LLMs -- e.g., \faComments \,\faHourglassO \textcolor{slightdarkgrey}{\textit{``I have become lazy [to write] even if [it is just a] small one line, I can also rely on the LLMs.''}} (IP16, R\&D Software Engineer). IP10 mentions trusting in LLMs for minor and low-risk tasks, which will not result in much damage, e.g., \faComments \textcolor{slightdarkgrey}{\textit{``If I am going to do some cosmetic changes, for example, [...] I just want to change how the page looks without actually affecting the logic. I think the LLMs will be a pretty safe bet to do because I'm not risking [inserting] bugs on the business logic''}} (IP10, Software Engineer).

\textbf{System Design \& Diagram Generation.} Two interview participants reported delegating system design or diagram generation to LLMs. IP12 collected the requirements and asked ChatGPT to generate the system design, providing the company context, e.g.: \faComments \,\faHourglass[3] \textcolor{slightdarkgrey}{\textit{``I let ChatGPT know: `These are the requirements, these are the stories that I collected from each person [...] Can you give me a good solution for this? And these are the existing systems that you can use [...] [According to ChatGPT,] this is just an authentication/authorisation system [...] Then I was like: `Yeah, this is good'''}} (IP12, Software Engineer). IP10 partially delegates diagram generation to LLMs: \faComments \,\faHourglass[3]  \textcolor{slightdarkgrey}{\textit{``If I have described an architecture, [I can say to the LLMs:] `Now, please create the diagram for this architecture'. [Then, the LLMs] it creates the mermaid diagram, so it can easily be visualised how this architecture works'''}} (IP10, Software Engineer). Related to diagram generation, Figure \ref{fig:workingMode} shows that 82\% of survey participants delegate code representation to LLMs, with 53\% delegating fully and 29\% partially.

 \textbf{Code Generation.} Five interview participants reported delegating code generation to LLMs. Figure \ref{fig:workingMode} shows that 70\% of survey participants partially delegate code generation to LLMs, and only 11\% fully delegate. Although IP20 still enjoys coding, they decide to delegate tedious or repetitive coding to LLMs: \faComments \,\faHourglass[3] \textcolor{slightdarkgrey}{\textit{``I still find sometimes writing code myself, it's way better than integrating with an LLM. And I only really use an LLM if the task that I'm about to do is really tedious and repetitive''}} (IP20, Software Engineer). Regarding tedious coding tasks, IP4 mentions the coding task of building a JSONBean: \faComments \,\faHourglassO \textcolor{slightdarkgrey}{\textit{``If you've ever done a query in JSONBean, the query gets very ugly. You write it about fifteen times until you get the syntax right. Then I said [to LLM]: `I have a JSON like this that has the information that's there, I created a SQL query for myself that I'm going to throw into the database, and it's going to return what I want'. That's it, the LLM generated it quickly and gave it to me.''}} (IP4, Software Developer).

\textbf{Code Translation.} Two interview participants reported delegating code translation to LLMs. For example, IP10 mentioned converting Python code to Golang code: \faComments  \,\faHourglassO \textcolor{slightdarkgrey}{\textit{``I had a working code in Python that I had to convert to Golang for another service. [...] because I already have all the codes done. It's already tested and everything. So I just need to do the conversion that might work for some things or others.''}}. Figure \ref{fig:workingMode} shows that 93\% of survey participants delegate code translation to LLMs, with 60\% delegating partially and 33\% fully.

\textbf{Test Generation.} Three interview participants reported delegating test generation to LLMs. For example, IP12 position test generation as a tedious low-stakes task: \faComments\,\faHourglassO \textcolor{slightdarkgrey}{\textit{``Because writing [tests] it is just very tedious. And I can miss some test cases, by the way. So, it's easy to give it to a machine algorithm to solve it rather than me doing that task. The downside of that is I still have to trust this system, whether the process still works. But if a unit test or integration test breaks, that's not the end of the world''}} (IP12, Software Engineer). SP74 uses LLMs to draft  tests and then refine them: \faComments\,\faHourglass[3] \textcolor{slightdarkgrey}{\textit{``I give the code to the LLM and let it write the whole test suite, then I review it and fix it if needed.''}} (Frontend Developer). 

% \textcolor{red}{survey results for test gen?}  @Samuel: Not included in the survey. 

\textbf{Document Generation.} Four interview participants mentioned delegating document generation to LLMs. For example, 
IP13 explains that they usually delegate to LLMs because of disliking writing code documentation: \faComments \,\faHourglassO   \textcolor{slightdarkgrey}{\textit{``[I use] mainly for documenting [...] it's just good [way] to save time, because no developer enjoys writing documentation, because when you're getting paid to code, you don't want to be writing documentation''}}. Figure \ref{fig:workingMode} shows that 90\% of survey participants delegate code comment generation to LLMs, with 48\% delegating partially and 42\% fully.

\textbf{Code Review.} Four interview participants mentioning delegating code review to LLMs. For example, IP13 also mentions that their company have incorporated AI-assisted pre-code review in their workflow: \faComments\,\faHourglass[3] \textcolor{slightdarkgrey}{\textit{``When I pull up a PR (pull request), we have an auto review bot that will come in and generate a [initial] review, but it won't approve the Pull Request. It will just generate [a code review with] all the strengths and weaknesses, and what exactly I'm fixing [in this PR] [...] And then the person who actually approves the PR does the [final] code review, [that person] can read through that as a summary''}} (IP13, Software Developer). Similarly, SP33 advocates for automated PR review to be the first defensive layer: \faComments\,\faHourglass[3] \textcolor{slightdarkgrey}{\textit{``You must not simply rely on the generated code to directly be pushed into production before careful review after the 1st level of automated PR reviews''}} (Software Engineer). 

% \textcolor{red}{survey results for code review?} @Samuel: Not included in the survey. 

\subsubsection{\textbf{{Full delegation vs Partial delegation}}}  \label{sec:qualitativeFindings} Our analysis of survey open-ended responses reveals that level of control, ownership and accountability, developer effort, developer mental model, and requirements play a role in developers' minds when comparing full delegation and partial delegation. For instance, developers need to closely manage the models' actions in partial delegation, e.g.: \faComments  \textcolor{slightdarkgrey}{\textit{``In partial delegation, there is continuous human oversight and the LLM is responsible for only specific tasks''}} (SP123, Software Engineer). On the other hand, minimal human input is required in full delegation, e.g.: \faComments   \textcolor{slightdarkgrey}{\textit{``Full delegation means handling off an entire task to the LLM with minimal human involvement''}} (SP118, Software Engineer).  In summary, participants differ partial delegation from full delegation in the level of human engagement, but both require human review.

% \textcolor{red}{anything else to add?}  

\subsubsection{\textbf{{Tasks developers self-delegate}}}

Only a few participants were interested in doing the following SE tasks mostly by themselves: code generation, test generation, pull request approval, and debugging. Their motivations for not delegating tasks to LLMs are mainly: false assurance, personal interests, unsatisfactory experience of debugging code using LLMs, and lack of trust in LLM tools.  

\textbf{Code Generation.} Only one interview participant showed a preference for coding themselves, without the support of LLMs. For instance, IP11  expressed distrust to LLMs' capabilites to work in complex coding generation tasks:  \faComments\,\faHourglass \textcolor{slightdarkgrey}{\textit{``When I was testing it in the early stage, I was trying to ask it to do something more complicated, like a little bit bigger stuff, [but] not [that] big, [the code generated via LLMs] will just have bugs''}}. Figure \ref{fig:workingMode} shows that 16\% of survey participants self-delegate code generation. \label{sec:codeGen}

\textbf{Test Generation.}  Two interview participants reported not using LLMs for testing.  IP19 defends that tests are vital for ensuring code quality, with no bugs: \faComments\,\faHourglass \textcolor{slightdarkgrey}{\textit{``So one of the things of test generation is: testing is the last thing that is actually holding the door between my code and some bug. So I should be very careful about using those tools there. So that's why - maybe - I didn't use that for test generation''}} (IP19, Machine Learning Scientist). SP125 prefers to review and test herself: \faComments\,\faHourglass \textcolor{slightdarkgrey}{\textit{``I ask LLMs to develop a login API which I review and test and harden before deploying it.''}} (DevOps Engineer). \label{sec:SelfDelegationTestGen}

\textbf{Pull Request Approval.} Only one interview participant defended that pull request approval should have human agency. IP9 \faComments \textcolor{slightdarkgrey}{\textit{``I'm testing pull request approval using LLMs. But I'm not fully confident because to do this, I [need to] rely more [on LLMs]''}} (AI Engineer).

\textbf{Debugging.} Two interview participants showed a preference for self-delegating debugging. IP11 reported having a bad experience using ChatGPT for debugging: \faComments \textcolor{slightdarkgrey}{\textit{``Asking GPT to debug is like [a] nightmare, and because it will just start to cycle on the problem''}} (Web Developer). Distrust in LLMs led SP145 to debug himself: \textcolor{slightdarkgrey}{\textit{``I believe that the current level of LLM is not that perfect; like, it's good for writing, but you have to shift to reading and verifying, and debugging code''}} (Software Developer).

\subsubsection{\textbf{{Task avoidance (despite LLMs availability)}}} \label{sec:qualitativeFindings}

Our findings indicate that certain developers deliberately resist AI assistance for specific tasks despite model availability. For instance, IP11 explicitly noted that they never write documentation even when they had access to LLMs. This stance aligns with our broader survey analysis, which identified a subset of participants who would prefer to avoid LLMs even when actively employing them for other development tasks. These tasks, illustrated in Figure \ref{fig:workingMode}, include code generation (n=4), code understanding (n=4), code completion (n=3), and data analysis (n=2). Respondents attributed this systematic aversion to several critical socio-technical barriers: perceived LLM limitations (e.g., the introduction of security vulnerabilities), strict requirements for handling sensitive information, the indispensable need for human review and legal accountability, a lack of developer expertise, a lack of trust in LLMs, specific task characteristics, and a fundamental desire to maintain human control over the codebase.

\begin{tcolorbox}
\textbf{RQ1 - Takeaway:} We identified four task delegation modes, or ways in which developers delegate tasks to LLMs: 1) \textit{full delegation}, (2) \textit{partial delegation}, (3) \textit{self-delegation}, and (4) \textit{task avoidance}.
\end{tcolorbox}

\subsection{Factors associated with Task Delegation (RQ2)}  \label{sec:RQ2}

\subsubsection{Emerging factors distilled from interviews} \label{factors} 
We identified task difficulty, task risk, motivation, and trust as potential factors driving LLM task delegation. 

\textbf{{Task Difficulty.}} This indicates the level of complexity, effort, and skill required for LLMs to complete the task without supervision. Some interview participants argue that they would delegate simple tasks to LLMs, e.g.: \faComments  \;\textcolor{slightdarkgrey}{\textit{``While [Microsoft Copilot] is really good to do some basic things that aren't mission critical, it struggles in a few areas where it still doesn't understand completely the task at hand''}} -- (IP14,  Data Engineer). However, LLMs are limited when it comes to complex task, e.g.: \faComments \textcolor{slightdarkgrey}{\textit{``When I think of a task, and how much time we would take, it also includes the complexity. So if I'm looking at a complex task, I know I probably wouldn't even be able to use ChatGPT on it.''}} -- (IP13, Software Engineer).

% They also highlight that LLMs are unsuitable to develop full software solutions, e.g.: \faComments  \textcolor{slightdarkgrey}{\textit{``We cannot have an agent that's developing everything on its own, because [...] SE is not only about writing code but it's also [about] designing and architecting the software as a whole.''}} -- (IP21, Research Engineer).  \textcolor{red}{first mention of agent -- do you have a different code about LLMs?}

\textbf{{Risk.}} This refers to any perceived individual or organisational risk from delegating tasks to LLMs, e.g., if the participant's organisation decided to prohibit LLMs because of potential risks: \faComments  \textcolor{slightdarkgrey}{ \textit{``They don't want to take any risk [...] if there is even like 0.1 chance of the code getting leaked to OpenAI, it's [considered] a big risk [for my company]''}} -- (IP11, Web Developer)

\textbf{{Motivation.}} This describes an developers' personal interest or enjoyment regarding a SE task. For instance, IP11 (Web Developer) reports enjoyment in coding and dislike for debugging and reading others' code: \faComments  \textcolor{slightdarkgrey}{\textit{``For me is that when you think about coding, the fun part is coding, and the worst part is debugging or reading others' code, especially debugging others' code''}}.

\begin{figure}[h]
    \centering
    \includegraphics[width=\linewidth]{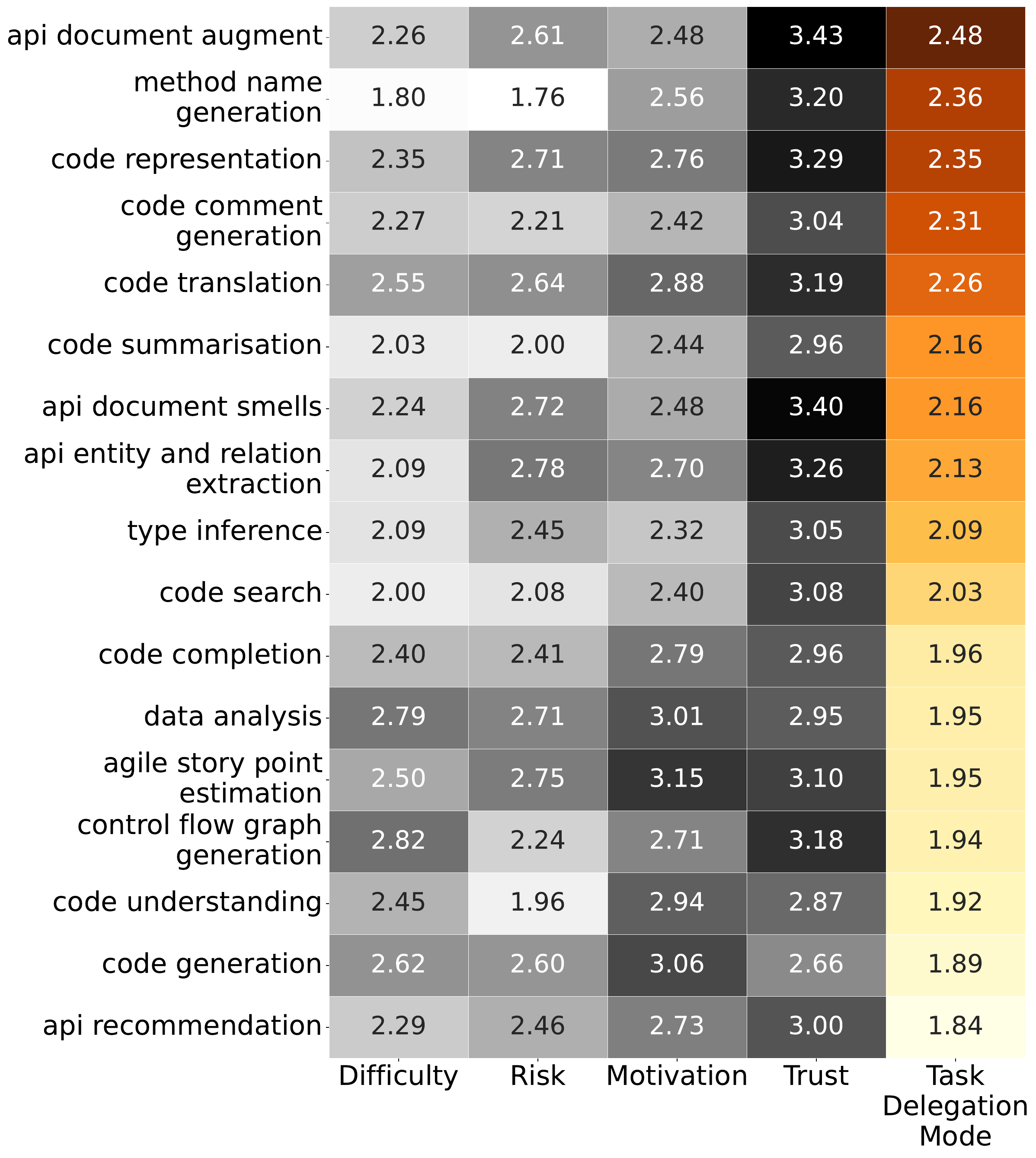}
    \caption{Emerging factors across tasks. \footnotesize{Task difficulty, perceived risk, developer motivation, and developer trust metrics vary from 0 to 4, while task delegation modes range from 0 to 3 (0: Task Avoidance; 1: Self-Delegation; 2: Partial Delegation; 3: Full Delegation)}.}
    \label{fig:TaskMetrics}
\end{figure}

\begin{table}
\centering
\caption{Sequential binary mixed-effects logistic regression
model results for (1) Adoption stage (N=857) and (2) Autonomy Gate (N=725). \footnotesize{Cells report standardised regression coefficients ($\beta$) and p-values.}}
\label{tab:hypothesesFindings}
\begin{tblr}{
  width = \linewidth,
  colspec = {Q[135]Q[317]Q[317]Q[171]},
  % column{4} = {font=\bfseries},
  cell{1}{1} = {font=\bfseries},
  hline{1,6} = {-}{0.08em},
  hline{2} = {-}{0.05em},
}
Factors         & {\textbf{Stage 1 }\\\textbf{ (No Del. vs. Del.)}} & {\textbf{Stage 2 }\\\textbf{ (Partial vs. Full)}} & \textbf{Empirical Finding}   \\
Difficulty ($H_1$) & {$\beta = -0.15$\\$p = 0.21$}                                    & {$\beta = -0.08$\\$p = 0.55$}                                    & Not Supported       \\
Risk ($H_2$)       & {$\beta = 0.15$\\$p = 0.18$}                                     & {$\beta = -0.20$\\$p = 0.17$}                                    & Not Supported       \\
Motivation ($H_3$) & {$\beta = -0.08$\\$p = 0.49$}                                    & {$\beta = 0.03$\\$p = 0.84$}                                     & Not Supported       \\
Trust ($H_4$)      & {$\mathbf{\beta = 0.70}$\\$\mathbf{p < 0.001^{***}}$}            & {$\mathbf{\beta = 1.30}$\\$\mathbf{p < 0.001^{***}}$}            & \textbf{Fully Supported}     
\end{tblr}
\smallskip
\flushleft{\scriptsize $^* p < 0.05$, $^{***} p < 0.001$. Statistically significant coefficients are highlighted in bold.}
\end{table}

\textbf{{Trust.}} This refers to the level of developer trust towards their LLMs. Interview participants mention different aspects of trust, from over-trust to distrust. For instance, one participant observed over-trust in LLMs when developers persist in vibe coding even after many unsuccessful interactions, e.g.: \faComments  \textcolor{slightdarkgrey}{\textit{``I have seen some of the junior developers who come in, they use the term vibe coding, and just like give the prompt and put it there. And sometimes, they just delete part of the codebase sometimes. And they're [persisting on LLMs, with the mindset]: `No, no, this is gonna work'''}} -- (IP12, Software Engineering).

\subsubsection{Descriptive results}  

Figure \ref{fig:TaskMetrics} captures the interplay between these four emerging factors  towards LLM task delegation. Each factor is measured on a 0–4 scale, and the resulting task delegation level is measured on a 0–3 scale across 17 coding-related tasks.  While the emerging factors fluctuate significantly - with \textit{difficulty} peaking at 2.82 (control flow graph generation), \textit{risk} at 2.78 (api entity and relation extraction), \textit{motivation} ranging from 2.32 to 3.15, and \textit{trust} hitting a low of 2.66 (code generation) - the task delegation scores are tightly bounded between 1.84 and 2.48. Because every mean score anchors closely to 2 (partial delegation), the data indicates that developers systematically reject the extremes of full delegation and no delegation (self-delegation or task avoidance), consistently opting for the balanced, partial delegation mode.  

% \textcolor{red}{Fig 5 is referred earlier in the text than Table IV but on the page Table IV appears first. Swap the placement of Fig 5 and Table IV so they appear in the right order?}

\subsubsection{Testing our Hypotheses}

Our sequential binary logistic regression model, shown in Table ~\ref{tab:hypothesesFindings} reveals that \textbf{trust is the only significant factor across the \textit{adoption} and \textit{autonomy} stages}. This provides a comprehensive overview of the estimated regression coefficients ($\beta$ in log-odds), their associated statistical significance, and the final verification status for each hypothesis across the two distinct decision stages. Task risk (\texttt{p=0.15; p=0.17}), task difficulty (\texttt{p=0.21; p=0.55}), and developer motivation (\texttt{p=0.49; p=0.84}) do not cross our strict significance threshold (\texttt{$\alpha$=0.05}). Thus, \textbf{our results do not support} $H_1$, $H_2$, or $H_3$. 

\noindent\textbf{Hypothesis 4:} \textit{Trust in LLMs is positively associated with delegation.} Our findings strongly support $H_4$, demonstrating that developer trust is a highly significant, positive factor across both  \textit{Adoption stage} (\texttt{log-odds $\beta$=0.70, p<0.001}) and the subsequent \textit{Autonomy stage} (\texttt{log-odds  $\beta$=1.30, p<0.001}). In Stage 1, a one-standard-deviation increase in trust was associated with $2.01$ times the odds of adoption (\texttt{OR=2.01;$R^2_m/R^2_c$=0.096/0.272;ICC=0.195}). Once the  decision to adopt an LLM has been made, the impact of trust significantly intensifies during the Autonomy stage (\texttt{OR=3.65;$R^2_m/R^2_c$=0.148/0.597;ICC=0.527}). Specifically, a one-standard-deviation increase in trust was associated with $3.65$ times the odds of full rather than partial delegation.

 % $H_1$ (\textit{task difficulty is negatively associated with delegation}), $H_2$ (\textit{task risk is negatively associated with delegation}), or $H_3$ (\textit{developer motivation is negatively associated with delegation}). 

% For every increase in trust - odds ratio (OR: 2.01) - developers have two times odds of delegating tasks to LLMs. 
\begin{tcolorbox}
\textbf{RQ2 - Takeaway:} \textit{Developer trust} is a significant factor associated with delegation (partially and fully). Trust shows a larger effect at the \textit{autonomy} stage (\texttt{OR:3.65}) -- transitioning from a \textit{partial} to \textit{full delegation} -- than at the \textit{adoption} stage (\texttt{OR:2.01}) -- transitioning from \textit{self-delegation} to \textit{partial} or \textit{full delegation}.
\end{tcolorbox}

% 

 % \textcolor{red}{consider including a super quick definition of autonomy stage and adoption stage in the brackets here. Drop the OR numbers if space is a problem and these numbers are mentioned somewhere else}

\section{Related Work}   \label{sec:relatedWork}

We summarise key related studies below, followed by a comparison of our findings to these works.

Through a survey with 188 software engineers, Lambiase et al. \cite{lambiase:2025} investigated the relationship between individual attributes related to technology adoption and LLM adoption for five specific SE tasks: manipulating artifacts, generating alternative versions of an artifact, retrieving information from external sources, supporting decisions through recommendations and contextual insights, and training in theoretical or practical knowledge of software concepts. They found that LLM adoption for SE might not be uniform across different tasks and argued that LLM adoption is a \textit{``domain-specific phenomenon rather than a broadly generalizable trend''}.

Sergeyuk et al. \cite{sergeyuk:2025} conducted a survey with 481 software developers to understand how software developers are using AI assistants, which tasks they are not using AI for, and what improvements are necessary in these assistants. They focused on five SE activities: implementing new features, writing tests, bug triaging, refactoring, and writing natural-language artifacts. Their findings indicate that: 84.2\% of the survey respondents occasionally or regularly used AI assistants; {implementing new features} was the most common activity; and code generation and code summarisation were the most common tasks. They also found that writing tests in general and generating data and resources for tests were less enjoyable tasks, which developers preferred to delegate to AI tools. They classified developers' motivations for not using LLMs into 20 themes. The most recurrent themes were the lack of need for AI assistance, AI-generated output being inaccurate, users' lack of trust and the desire to feel in control, and limitations of AI assistants in understanding context. Our findings align with Sergeyuk \cite{sergeyuk:2025} regarding the relationship between perceived task difficulty and willingness to delegate. Their correlation analysis across five SE tasks similarly demonstrated a lack of statistical significance. However, our findings arise from evaluating these dynamics across a larger pool of SE tasks and based on tasks that our participants reported having experience in using LLMs.

Choudhuri et al. \cite{choudhuri:2025} explored how software developers' evaluations of their tasks shape their preference for where, why, and how AI should support their workflows. They conducted a survey with 840 Microsoft developers using perceived task significance (\textit{value}), level of alignment with developers' interests and expertise (\textit{identity}), perceived responsibility and blame (\textit{accountability}), and task effort (\textit{demands}) as study metrics. They used their synthesised literature-grounded taxonomy of SE tasks. They also employed Microsoft's Responsible AI framework \cite{microsoft:2026} to assess developers' Responsible AI priorities. Their findings indicate that task value, accountability, and demands positively influence AI openness and use, while developers' interests and expertise (identity) may lead to lower public support for AI, even as their usage remains higher. They also observed great receptivity from junior, AI-experienced, risk-tolerant, and technophilic developers. Our results reinforce those of Choudhuri et al. \cite{choudhuri2:2025}, who found that trust is directly associated with actual LLM usage among software developers at Microsoft and GitHub. Our study expands on their work by demonstrating that trust is vital not just for adoption, but specifically for navigating both the initial adoption and subsequent autonomy stage.

% Our findings only partially confirm the experimental results reported by Spitzer et al. \cite{spitzer:2025}. Their regression analysis indicated that perceived difficulty increases the likelihood of delegating tasks to AI. In contrast, while our qualitative interview data suggested a similar trend, our quantitative survey analysis found no statistically significant support for this. This discrepancy may be due to task differences: Spitzer et al. evaluated delegation behaviour using an income prediction task \textcolor{red}{what can income prediction mean? was this even about SE?} \textcolor{blue}{Samuel: It is not specific about SE but task delegation (human-AI collaboration)}, whereas our study investigates delegation across software development tasks. \textcolor{red}{May not be a closely related work - drop this? Compare to Lambiase instead?}

% \textcolor{red}{no comparison to Lambiase et al.?}

% \textcolor{red}{since there are only 3-4 key RWs, you can present one study summary followed by how our findings compare to that study, then the next study summary and comparison and so on}

\section{Discussion}  \label{sec:discussion}

\subsection{Implications for Practice}
Based on our findings and corroborating with related works, we suggest the following implications for software practitioners.\\

%\begin{itemize}
% \textcolor{red}{YIKES! Move RWs to before Discussion so this makes sense. It will still be in one place as opposed to two and toward the end - then double check reference/sequence is fine elsewhere, e.g., at the end of the intro}

    \noindent \faChevronCircleRight ~\textbf{Do not expect developers to blindly delegate code generation to LLMs.} Figure~\ref{fig:workingMode} shows that 70\% of the participants partially delegate, 16\% self-delegate, and only 11\% fully delegate \textit{code generation}. The level of trust plays a significant role in this decision. Figure~\ref{fig:TaskMetrics} shows that \textit{code generation} ranks as the least trusted task (scoring 2.66 out of 4). The potential bugs from AI-generated code seems to make developers distrust LLMs for code generation (Section \ref{sec:codeGen}).
    
    % , e.g.:   \faComments\textcolor{slightdarkgrey}{\textit{``When I was testing it [...] I was trying to ask it to do something more complicated [...] [the code generated via LLMs] will just have bugs''}} (IP11, Web Developer). \textcolor{red}{is this quote repeated? doesn't seem to add much here anyway. remove?}

    \noindent \faChevronCircleRight ~\textbf{Developers select partial delegation for most tasks -- reflecting a balanced control towards LLMs.}  Partial delegation is the most frequent task delegation mode across the analysed tasks. Our findings show that 60\% of developers choose this mode even though it demands more engagement from the developer as compared to full delegation (see Figure \ref{fig:workingMode}). This prevalent choice reflects a deliberate effort by developers to keep the code under active human control rather than handing it over blindly to the model, an idea captured as a \textit{balanced control} in the reliance-control framework \cite{ferino2:2026}.

    \noindent \faChevronCircleRight ~\textbf{Team task risk alignment is necessary.} 
    Our findings show that developers perceive task risk varying according to different tasks (see Figure \ref{fig:TaskMetrics}). The danger is that developers undermine crucial tasks. For instance, as highlighted in Section \ref{sec:SelfDelegationTestGen}, some developers may misclassify tests as a low-stakes task. In this scenario, the team should align on how much they are willing to delegate to LLMs, balancing productivity gains and reliability. 
    
    % \textcolor{red}{good point - reminds me that team wide task delegation can be an idea for future research}

    \noindent \faChevronCircleRight ~\textbf{Trust is a significant factor associated with the level of task delegation to LLMs.} Through our mixed methods study, we identified \textit{trust} as a significant factor associated with task delegation (Table \ref{tab:hypothesesFindings} and Figure \ref{fig:TaskMetrics}). Our study provides empirical evidence of the claim made in the Communications of the ACM opinion piece by Roychoudhury et al. that said \textit{``with the increased use of AI code generation [...] the emphasis will be not only on programming at scale, but increasingly on programming with trust''} \cite{roychoudhury:2026}, and highlights the need to better understand and improve developer trust.

    \noindent \faChevronCircleRight ~\textbf{From delegation to automation -- LLM-based tool developers should focus on features that improve developer trust.} Since trust is a significant factor associated with partial or full dev-LLM task delegation, LLM-based software development tools need features that help improve developer trust. For example, with increased developer trust, tasks close to full delegation (Fig. \ref{fig:TaskMetrics}) could progress toward automation, reaching what has been referred to as \textit{escape velocity} in industry, described as \textit{``the shift from AI as a tool to AI as an autonomous collaborator''} \cite{subbu:2026}. Hassan et al. \cite{hassan:2025} capture this notion in \textit{specialised domain autonomy} (level 4) and \textit{general domain autonomy} (level 5) of their hierarchical framework for AI in SE. Efforts such as Wang et al. \cite{wang3:2024}, who designed a dashboard to communicate aggregate GitHub Copilot usage, represent initial steps toward designing for trust in LLM tools.

\subsection{Implications for Research} %include ideas for future works
Based on our study, we share some implications for research and ideas for future work.\\

        \noindent  \faChevronCircleRight ~\textbf{What other factors drive task delegation?} Future research could investigate other socio-technical factors, such as \textit{application domain} (e.g., safety critical domains such a healthcare vs entertainment domains such as gaming) and \textit{interaction modes} (e.g., chat-mode, voice mode, agentic terminal, IDE based AI). A deeper dive into how \textit{trust} is manifested, enabled, and inhibited in developer-LLM collaboration will also help. Khati et al.  \cite{khati:2025} found that developer trust is associated  with many factors (e.g., accuracy, interpretability) in their systematic literature review. They also pointed out that LLM trustworthiness, not just trust, must be addressed.

        \noindent \faChevronCircleRight ~\textbf{How to deal with the code review bottleneck?} Our findings indicate that even during full delegation, participants tend to review the LLM output. DORA report \cite{dora:2025} warned that the velocity gains for code generation may result in massive blocks of code, which may take a longer time for developers to review and test. Researchers could explore new visualisation tools designed specifically to help developers identify logic flaws or regressions in large-scale LLM outputs. 

        \noindent \faChevronCircleRight ~\textbf{What is the impact of organisational policies?} Our demographics data show a wide variation in company AI policies (ranging from allowing to strictly prohibiting/restricting), where participants engaged with LLMs in varying capacities regardless. Future research can study how institutional surveillance, corporate liability frameworks, and formal restrictions influence developer trust and actual task delegation modes in daily work.
        
        \noindent \faChevronCircleRight ~\textbf{Human-AI collaboration research.}  Our experience in this study suggests that studying human-AI collaboration topics can be better addressed by explicitly intersecting three core dimensions: the \textit{human} role (e.g., developer, tester, end-user), the \textit{AI} specifics (e.g., LLM, agents, multi-agent systems), and the \textit{collaboration}: practice/activity (e.g., task delegation), evolving role (e.g., project manager), or artefact (e.g., documentation). Future researchers can leverage this triad to design contextualised and in-depth empirical studies. 
% Rather than addressing LLMs as a monolithic tool,

\section{Threats to Validity} \label{sec:threatsToValidity}

In this section, we discuss the limitations and threats to the validity of this study. We employ a combined approach \cite{storey:2025} to discussing the study limitation for the applied qualitative methods \cite{korstjens:2018} and quantitative methods \cite{wohlin:2012}.

\textit{Construct Validity \& Confirmability} -- assesses how effectively the methodology measures the phenomenon that was intended to be captured. The task delegation modes and hypotheses were derived from our exploratory interview findings. We leveraged the TDAI framework \cite{lubars:2019, cvetkovic:2022} to investigate how risk, difficulty, motivation, and trust are associated with SE task delegation to LLMs. We acknowledge our findings are limited to self-reported data and thus, subject to self-reporting bias.

\textit{Internal Validity \& Credibility} -- assesses whether empirical evidence robustly supports the study findings. We acknowledge that selection bias is an inherent threat in our study, as participants voluntarily opted to join both the interviews and the survey. We mitigate this by selecting only software practitioners with experience using LLMs for software development. We also recognise that research bias may have occurred during our qualitative and quantitative analysis. To ensure credibility, the first author conducted data analysis pilots in both interview and survey data analysis and was supervised by qualitative and mixed research experts. The first author also consulted two expert statisticians for guidance on the most suitable statistical approach.
We observed considerable variation in the number of responses across tasks. Because we did not measure how frequently participants performed each task daily (to avoid respondent attrition due to an over lengthy survey), some differences in delegation patterns may reflect differences in participants' exposure to particular tasks rather than task characteristics alone.

\textit{Conclusion Validity \& Dependability} -- evaluates to which extension the conclusions can be reached from the data collected. Although our human ethics approval \textbf{(\#xxxx - withheld for anonymous reviewing)} does not allow us to share the interview and survey raw data obtained in this study, we adopted some strategies to transparently document and communicate the data analysis process. For example, we provide an example of open coding, constant comparison, and memoing (see Figure \ref{fig:OpenCodeConstantComparison}). We also provide plenty of original quotes throughout the paper. Finally, we provide the research artifacts \cite{researchArtifact},  scripts and spreadsheets, created during the data analysis process.

% improve the generalizability

\textit{External Validity \& Transferability} -- examines the extent to which our study findings can be generalised. Seeking to expand the scope of our interview findings,  we conducted our global survey study. However, to ensure the survey remained concise and doable, we narrowed the scope of tasks from the entire SDLC to 17 specific coding tasks. Our decision was based on the feedback received from five rounds of pilot surveys. Both our interview and survey participants encompass a diverse range of professional experiences, geographical locations, and industry domains lending reasonable stretch value to the findings in similar contexts.

\section{Conclusion}   \label{sec:conclusion}
Our  mixed-methods study set out to understand how software developers delegate tasks to LLMs in real-world practice, to what extent, and why. We interviewed 22 practitioners and surveyed 160 practitioners. Based on the analysis, we identified four software task delegation modes:  \textit{full delegation} to LLMs,  \textit{partial delegation} to LLMs, \textit{self-delegation}, and \textit{task avoidance}. Understanding these four task delegation modes is critical for both software engineering research and practice. For software organizations, these modes provide a concrete vocabulary to audit delegation patterns and evaluate risk exposure. For the research community, this study opens new avenues for modeling the socio-technical factors governing human-AI collaboration.

We also explored why developers delegate the tasks they do and identified how \textit{task difficulty}, \textit{task risk}, \textit{developer motivation}, and \textit{developer trust towards LLMs} are associated with task delegation.  Our findings confirm that \textit{developer trust} towards LLMs is a significant factor associated with task delegation. For software organisations aiming to move developer-AI interactions from basic AI assistance toward autonomous task delegation,  building a balanced level of developer trust in LLMs’ capabilities is essential.

% , not only improving trustworthiness in LLMs but also  building trust in LLMs' capabilities (e.g., accuracy, interpretability) is essential.

% trustworthiness in LLMs and developer trust towards LLMs are a priority. 

% building trust in LLMs' capabilities through (e.g., accuracy, interpretability)

% \textcolor{red}{not only LLM capabilities, there were other aspects to trust, right? list here} is essential. 

% LLMs trustworthiness s
% Without improving trust, developers will intentionally limit their LLM usage to partial delegation, prematuring capping potential gains -- no evidence to support this conclusion!

% LLM adoption -- strange to bring up LLM adoption here in the conclusion! drop. focus on task delegation.
%  in developer experience -- no evidence of this concept in the study. better not bring up right at the end!

\section{Acknowledgment}

Samuel is supported by a Faculty of IT PhD scholarship. We would like to express profound gratitude to all participants who took part in this study, Humphrey Obie for the wonderful feedback on this research, and Nimmi Weeraddana and Haoyu Gao for helping during participant recruitment. This research was supported by the Monash Statistical Consulting Service, part of Monash eResearch.

% https://dl.acm.org/doi/pdf/10.1145/3377811.3380387
% https://dl.acm.org/doi/pdf/10.1145/3757060
\bibliographystyle{IEEEtran}
\bibliography{IEEEabrv,references}

\end{document}